\documentclass[11pt]{article}
\usepackage{amsmath,amsfonts,amsthm,mathrsfs}

\newcommand{\R}{{\mathbb R}} \newcommand{\N}{{\mathbb N}}

\newcommand{\Prm}{{\mathbb P}}

\renewcommand{\epsilon}{\varepsilon }

\renewcommand{\rho}{\varrho } 
 
\renewcommand{\phi}{\varphi }

\newcommand{\q}{{\rm q }}

\newtheorem{lemma}{Lemma}
\newtheorem{corollary}{Corollary}
\newtheorem{proposition}{Proposition}

\begin {document}
 \title{Quantum Lower Bounds by Entropy Numbers}
\date{}
\author {Stefan Heinrich\\
Department of Computer Science\\
University of Kaiserslautern\\
D-67653 Kaiserslautern, Germany\\
e-mail: heinrich@informatik.uni-kl.de\\
homepage: http://www.uni-kl.de/AG-Heinrich}   
\date{}
\maketitle
\begin{abstract} 
We use entropy numbers  in combination with the polynomial method to 
derive a new general lower bound for the $n$-th minimal error in the quantum setting of information-based complexity. 
As an application, we improve some lower bounds on quantum approximation of embeddings between finite dimensional $L_p$ spaces and
of  Sobolev embeddings. 
\end{abstract}

\section{Introduction}
There is one major technique for proving lower bounds in the
quantum setting of IBC (information-based complexity) as introduced in \cite{Hei01}. 
It uses the polynomial  
method  \cite{BBC98} together with a result
on approximation by polynomials from \cite{NW99}. This method has been applied in \cite{Hei01},\cite{HN01b},\cite{Wie04}.
Other papers on the quantum complexity of continuous problems use
this implicitly by reducing mean computation to the problem in consideration and then using the lower bound for mean computation
of \cite{NW99} directly (\cite{Nov01},\cite{TW01},\cite{Kac05}, \cite{PW05}). 

This approach, however, does not work for the case of approximation of embedding operators in spaces with norms different from the infinity norm.
To settle such situations, a more sophisticated way of reduction to known bounds was developed in \cite{Hei04a}, based on a 
multiplicativity property of the $n$-th minimal quantum error. 

In this paper we introduce an approach which is new for the IBC quantum setting. 
We again use the polynomial method of \cite{BBC98}, but combine it with methods related to 
 entropy \cite{CS90}. We 
derive lower bounds for the $n$-th minimal quantum error in terms of certain entropy numbers.
Similar ideas have been applied before in \cite{Shi02}, the model and methods however being different, see also related work \cite{Nay99}.
As an application, 
we improve the lower bounds \cite{Hei04a,Hei04b} on approximation as well as those of \cite{Hei04c} by 
removing the logarithmic factors. 

 Let us also mention that a modification of the polynomial method based on trigonometric polynomials was used  in \cite{Bes04,Bes05}
 for proving  lower bounds for a type of query different from that introduced in \cite{Hei01},
the so-called power-query \cite{PW05}. 
 Our method can also be applied in this setting and simplifies the analysis from \cite{Bes04,Bes05}. 
 We comment on this at the end of the paper. 
 
\section{Lower bounds by entropy}

We work in the quantum setting of IBC as introduced in \cite{Hei01}. We refer to this paper for the needed notions.
Let $D$ and $K$ be nonempty sets, let $\mathcal{F}(D,K)$ denote the set of all functions on $D$ with values in $K$,
let $F\subseteq \mathcal{F}(D,K)$ be nonempty, and let $G$ be a normed linear space.
Let $A$ be a quantum algorithm from $F$ to $G$. For any subset $C\subseteq G$ define the function
$p_C:F\to \R$ by 
$$
p_C(f)=\Prm\{A(f)\in C\}\qquad (f\in F)
$$
-- the probability that the output of algorithm $A$ at input $f$ belongs to $C$. This quantity is well-defined for all subsets $C$ since the output of $A$ takes only finitely many values,
see \cite{Hei01}.
Furthermore, define 
$$
\mathcal{P}_{A,F}={\rm span}\{p_C :\, C\subseteq G\}\subseteq \mathcal{F}(F,\R)
$$
to be the linear span of the functions $p_C$. 

We need some notions related to entropy. We refer to \cite{CS90} for the definitions.
For a nonempty subset $W$ of a normed space $G$ and  $k\in \N$ (we use the notation $\N=\{1,2,\dots\}$ and
$\N_0=\{0,1,2,\dots\}$) define the $k$-th inner entropy number as
\begin{eqnarray}
\label{BD2}
\phi_k(W,G)&=&\sup\{\epsilon\,:\,\text{there exist } u_1,\dots,u_{k+1}\in W \text{ such that }\nonumber\\
&&  \|u_i-u_j\|\ge 2\epsilon \; \text{ for all }\; 1\le i\ne j\le k+1\}.\label{BD1}
\end{eqnarray}
It is worth while mentioning a related notion. The $k$-th entropy number is defined to be 
\begin{eqnarray}
\epsilon_k(W,G)&=&\inf\big\{\epsilon\,:\,\text{there exist } g_1,\dots,g_{k}\in G \text{ such that }\nonumber\\
&&\qquad\min_{1\le i\le k}\|g-g_i\|_G\le \epsilon \;\text{ for all  }\; g\in W \big\}.\label{BB3}
\end{eqnarray}
Then
\begin{equation}
\label{BD3}
\phi_k(W,G)\le \epsilon_k(W,G)\le 2 \phi_k(W,G),
\end{equation}
see \cite{CS90}, relations (1.1.3) and (1.1.4). Also  observe that the first numbers of both types are related to 
the radius and diameter of $W$ as follows:
\begin{equation}
\label{BG5}
\phi_1(W,G) =\frac12\, {\rm diam} (W,G),\quad \epsilon_1(W,G)={\rm rad} (W,G).
\end{equation}
Entropy numbers of bounded linear operators (that means, the 
entropy numbers of the image of the unit ball under 
the action of the operator) as well 
as their relation to various $s$-numbers and to eigenvalues are well-studied,
see again \cite{CS90} and references therein.

Let $S$ be a mapping from $F$ to $G$ 
and let $e(S,A,F)$ denote the error of quantum algorithm $A$ on $F$.
Our basic lemma relates this error to the dimension of $\mathcal{P}_{A,F}$ and the entropy of
$S(F)\subseteq G$.

\begin{lemma}
\label{lem:B1}  
(i) Let 
 $k\in \N$ be such that
\begin{equation}
\label{BF1}
k+1>(\log_2 5)\dim \mathcal{P}_{A,F}.
 \end{equation}
Then
\begin{equation}
\label{BF2}
e(S,A,F)\ge\phi_k(S(F),G).
\end{equation}
(ii) If $A$ is an algorithm without queries, then
\begin{equation}
\label{BG1}
e(S,A,F)\ge\phi_1(S(F),G).
\end{equation}
\end{lemma}
\begin{proof} The first part of the proof is the same for both cases. For case (i) we assume that $k$ satisfies 
(\ref{BF1}), while in case (ii) we set $k=1$. 
Let $f_1,\dots,f_{k+1}\in F$ be arbitrary elements and put
\begin{equation}
\label{BA2}
\epsilon=\min\big\{ \|S(f_i)-S(f_j)\| \,:\, 1\le i\ne j\le k+1 \big\}.
\end{equation}
It sufffices to show that 
\begin{equation}
\label{BB1}
e(S,A,F)\ge \epsilon/2.
\end{equation}
For $\epsilon =0$ this is trivial, so we suppose $\epsilon >0$. We assume the contrary of (\ref{BB1}), that is
\begin{equation}
\label{BB2}
e(S,A,F)<\epsilon/2.
\end{equation}
By (\ref{BA2}), the subsets $V_i\subset G$ defined by 
\begin{equation}
\label{BA8}
V_i=\left\{g\in G\,:\, \|S(f_i)-g\|<\frac{\epsilon}{2}\right\} \quad(i=1,\dots,{k+1})
\end{equation}
are disjoint.
It follows from (\ref{BB2}) and (\ref{BA8}) that for $i=1,\dots,k+1$
\begin{equation}
\label{BG2}
\Prm\{A(f_i)\in V_i\}\ge \frac34.
\end{equation}
Let us first complete the proof of (ii): If $A$ has no queries, its output does not depend on $f\in F$, and in particular, the
distribution of the random variables $A(f_1)$ and $A(f_2)$ is the same. But then (\ref{BG2}) implies
$$
\Prm\{A(f_1)\in V_1\cap V_2\}\ge 1/2,
$$
 thus $  V_1\cap V_2\ne \emptyset$, a contradiction, which proves (\ref{BB1}) in case (ii).

 Now we deal with case (i). Let $\mathcal{C}$ be the set of all $C\subset G$ of the form 
$$
C=\bigcup_{i\in I} V_i,
$$
with $I$ being any subset of $\{1,\dots, k+1\}$. Clearly, 
\begin{equation}
\label{BA3}
|\mathcal{C}|=2^{k+1}.
\end{equation}
Let $\mathcal{P}_{A,F}$ be endowed with the supremum norm 
$$
\|p\|_\infty=\sup_{f\in F} |p(f)|.
$$
We have 
\begin{equation}
\label{BA6}
\|p_C\|_\infty\le 1 \quad (C\in \mathcal{C}). 
\end{equation}
Moreover,
\begin{equation}
\label{BA7}
\|p_{C_1}-p_{C_2}\|_\infty\ge \frac12\quad (C_1\ne C_2\in\mathcal{C}).
\end{equation}
Indeed, for $ C_1\ne C_2\in\mathcal{C}$ there is an $i$ with $1\le i\le {k+1}$ such that
$V_i \subseteq C_1\setminus C_2$ or $V_i \subseteq C_2\setminus C_1$. Without loss of generality we assume the first.
Then, because of (\ref{BG2}), we have
$$
p_{C_1}(f_i)=\Prm\{A(f_i)\in C_1\} \ge \Prm\{A(f_i)\in V_i\}\ge \frac34,
$$
while 
$$
p_{C_2}(f_i)=\Prm\{A(f_i)\in C_2\}\le\Prm\{A(f_i)\in G\setminus V_i\}\le \frac14,
$$
hence
$$
|p_{C_1}(f_i)- p_{C_2}(f_i)|\ge \frac12,
$$
implying (\ref{BA7}). For $p\in \mathcal{P}_{A,F}$ let $B(p,r)$ be the closed ball of radius $r$ around $p$ in  $\mathcal{P}_{A,F}$.
By (\ref{BA7}) the balls $B(p_C,1/4)$ have disjoint interior for $C\in \mathcal{C}$. Moreover, by (\ref{BA6}), 
$$
\bigcup_{C\in \mathcal{C}}B(p_C,1/4)\subseteq B(0,5/4).
$$ 
A volume comparison gives 
$$
2^{k+1}=|\mathcal{C}|\le 5^{\dim \mathcal{P}_{A,F}},
$$
hence, taking logarithms, we get
a contradiction to (\ref{BF1}), which completes the proof.
\end{proof}
Let $e_n^\q(S,F)$ denote the $n$-th minimal quantum error, that is, the infimum of $e(S,A,F)$ taken over
all quantum algorithms $A$ from $F$ to $G$ 
with at most $n$ queries (see \cite{Hei01}). As an immediate consequence of Lemma \ref{lem:B1}, and also for later use, we note
the following.
\begin{corollary}
\label{cor:B1}
\begin{equation}
\label{BG3}
\frac12\,{\rm diam}(S(F),G)\le e_0^\q(S,F)\le {\rm rad}(S(F),G).
\end{equation}
\end{corollary}
\begin{proof}
The lower bound follows from  Lemma \ref{lem:B1} (ii) and (\ref{BG5}). The upper bound is obtained by taking for any $\delta>0$ 
a point $g_\delta\in G$ with 
$$
\|S(f)-g_\delta\|\le {\rm rad}(S(F),G)+\delta\quad \text{for all }\; f\in F
$$
and then using the trivial algorithm which outputs $g_\delta$ for all $f\in F$, with probability 1. 
\end{proof}
Next we recall some facts from \cite{Hei01}, section 4.
Let
$L\in \N$ and let to each $u=(u_1,\dots,u_{L})\in\{0,1\}^L$ an 
$f_u\in \mathcal{F}(D,K)$ be assigned such that the following 
is satisfied:
\\ \\
{\bf Condition (I):} For each $t\in D$ there is an $\ell$, $1\le \ell\le L$, such that $f_u(t)$ 
depends only on $u_\ell$, in other words, for $u,u'\in\{0,1\}^L$, 
$u_\ell=u'_\ell$ implies $f_u(t)=f_{u'}(t)$.
\\ \\ 
The following result was shown in \cite{Hei01}, Corollary 2, based on the idea of the quantum polynomial method
 \cite{BBC98}.
\begin{lemma}
\label{lem:B2}
Let $L\in\N$ and assume that  $(f_u)_{u\in\{0,1\}^L}\subseteq\mathcal{F}(D,K)$
satisfies condition (I). 
Let $n\in \N_0$ and let $A$ be a quantum algorithm from $\mathcal{F}(D,K)$ to $G$ with $n$ quantum queries.
Then for each subset $C\subseteq G$, 
$$
p_C(f_u)=p_C\left(f_{(u_1,\dots,u_L)}\right),
$$
considered as a function of the variables $u_1,\dots,u_{L}\in\{0,1\}$, is a real multilinear polynomial of degree at most $2n$.
\end{lemma}
Now we are ready to state the new lower bound on the $n$-th minimal quantum error.
\begin{proposition}
\label{pro:B1} 
Let $D,K$ be nonempty sets, let 
$F\subseteq\mathcal{F}(D,K)$ be a nonempty set of functions, $G$ a normed space,
$S:F\to G$ a mapping, and $L\in\N$. Suppose 
$\mathcal{L}=(f_u)_{u\in\{0,1\}^L}\subseteq\mathcal{F}(D,K)$ is a system of functions satisfying
condition (I). Then
\begin{equation*}
e_n^\q(S,F)\ge \phi_k(S(F\cap \mathcal{L}),G)
\end{equation*}
whenever $k,n\in \N$ satisfy $2n\le L$ and 
\begin{equation}
\label{BF5}
k+1>(\log_2 5) \left(\frac{eL}{2n}\right)^{2n}.
\end{equation}
\end{proposition}
\begin{proof}Let $n\in \N$ with $2n\le L$  and let $A$ 
be a quantum algorithm  from $F$ to $G$ 
with no more than $n$ queries. Note that, by definition, a quantum algorithm from $F\subseteq \mathcal{F}(D,K)$ to $G$
is always also a quantum algorithm from  $\mathcal{F}(D,K)$ to $G$ (see \cite{Hei01}, p.\ 7). We show that
\begin{equation}
\label{BF3}
e(S,A,F)\ge \phi_k(S(F\cap \mathcal{L}),G)
\end{equation}
for all $k\in \N$ satisfying (\ref{BF5}).
Let $\mathscr{M}_{L,2n}$ be the linear space of real multilinear polynomials in 
$L$ variables of degree not exceeding $2n$. Since $2n\le L$, its dimension is
\begin{equation}
\label{BA9}
\dim \mathscr{M}_{L,2n}=\sum_{i=0}^{2n} {L \choose i} \le \left(\frac{eL}{2n}\right)^{2n}
\end{equation}
(see, e.g., \cite{Mat99}, (4.7) on p.\ 122, for the inequality). Set
$$
U=\{u\in\{0,1\}^L:\,f_u\in F\}
$$ 
and let $\mathscr{M}_{L,2n}(U)$ denote 
the space of all restrictions of functions from $\mathscr{M}_{L,2n}$ to $U$.
Clearly,
\begin{equation}
\label{BF6}
\dim \mathscr{M}_{L,2n}(U)\le \dim \mathscr{M}_{L,2n}.
\end{equation}
Define 
$$
\Psi:\mathcal{P}_{A, F\cap\mathcal{L}}\to \mathcal{F}(U,\R)
$$
by setting for $p\in\mathcal{P}_{A, F\cap\mathcal{L}}$ and $u\in U$
$$
(\Psi p)(u)=p(f_u).
$$
Obviously, $\Psi$ is linear, moreover, for $C\subseteq G$
$$
(\Psi p_C)(u)=p_C(f_u)\quad(u\in U).
$$
By Lemma \ref{lem:B2}, $p_C(f_u)$, as a function of $u\in U$, is the restriction of an element of  $\mathscr{M}_{L,2n}$ to $U$.
Hence,
$$
\Psi p_C\in \mathscr{M}_{L,2n}(U),
$$
and by linearity and the definition of $\mathcal{P}_{A, F\cap\mathcal{L}}$ as the linear span of functions $p_C$,
we get
$$
\Psi(\mathcal{P}_{A, F\cap\mathcal{L}})\subseteq \mathscr{M}_{L,2n}(U).
$$
Furthermore, $\Psi$ is one-to-one, since $\{f_u:\,u\in U\}=F\cap\mathcal{L}$. Using (\ref{BA9}) and (\ref{BF6}) it follows that
$$
\dim\mathcal{P}_{A, F\cap\mathcal{L}}\le \dim \mathscr{M}_{L,2n}(U)\le \left(\frac{eL}{2n}\right)^{2n}.
$$
Consequently, for $k$ satisfying (\ref{BF5}),
$$
k+1>(\log_25)\dim\mathcal{P}_{A, F\cap\mathcal{L}}.
$$ 
Now (\ref{BF3}) follows from Lemma \ref{lem:B1}.
\end{proof}

\section{Some applications}
For $N\in\N$ and  $1\le p\le\infty$, 
let $L_p^N$ denote the space of all functions
$f:\{1,\dots,N\}\to \R$, equipped with the norm 
$$
\|f\|_{L_p^N}=\left(\frac{1}{N}\sum_{i=1}^{N}|f(i)|^p \right)^{1/p}
$$
if $p<\infty$,
$$
\|f\|_{L_\infty^N}=\max_{1\le i\le N} |f(i)|,
$$
 and let $\mathcal{B}(L_p^N)$ be its unit ball. Define $J^N_{pq}:L_p^N\to L_q^N$ to be the identity operator 
$J^N_{pq}f=f\;\,(f\in L_p^N)$.

As already mentioned, the
lower bound for approximation of $J^N_{pq}$  
was obtained using a multiplicativity property of the $n$-th minimal quantum error (\cite{Hei04a}, Proposition 1).
The result involved some logarithmic factors of negative power (\cite{Hei04a}, Proposition 6). 
Based on Proposition \ref{pro:B1} above we improve this bound by removing the logarithmic factors.

\begin{proposition}
\label{pro:B2} 
Let $1\le p,q\le  \infty$. There is a constant $c>0$ such that 
for all $n\in \N_0$, $N\in\N$ with  $n\le c N$ 
$$
e_n^\q(J_{p,q}^{N},\mathcal{B}(L_p^N))\ge \frac18.
$$
\end{proposition}
\begin{proof}
It suffices to prove the case $p=\infty$, $q=1$. 
We put $L=N$ and $f_u=u$ for $u\in \{0,1\}^N$. Clearly, the system 
$\mathcal{L}=(f_u)_{u\in \{0,1\}^N}$ satisfies condition (I) and
\begin{equation}
\label{BG7}
\mathcal{L}\subset \mathcal{B}(L_\infty^N).
\end{equation}
Let $\{ f_{u_i}\,:\, 1\le i\le k+1\}$ be a maximal system with 
\begin{equation}
\label{BH2}
\|f_{u_i}-f_{u_j}\|_{L_1^N}\ge \frac14\qquad (1\le i\ne j\le k+1).
\end{equation}
Maximality implies 
$$
\{0,1\}^N= \bigcup_{i=1}^{k+1}\left\{ u\in\{0,1\}^N\,:\, \|f_u-f_{u_i}\|_{L_1^N}<\frac14\right\}.
$$
On the other hand, 
\begin{eqnarray*}
2^N&\le& \sum_{i=1}^{k+1}\left|\left\{ u\in\{0,1\}^N\,:\, \|f_u-f_{u_i}\|_{L_1^N}<\frac14\right\}\right|\\
&\le& (k+1)\sum_{0\le j<N/4}{N\choose j}\le (4e)^{N/4 },
\end{eqnarray*}
again by \cite{Mat99}, (4.7) on p.\ 122. It follows that 
\begin{equation}
\label{BF7}
k+1\ge 2^N (4e)^{-N/4}=2^{c_1N},
\end{equation}
with $c_1=\frac14\log_2(\frac4e)>0$, hence $k\in\N$.
From  (\ref{BH2}) we obtain 
\begin{equation}
\label{BF8}
\phi_k\left(J_{\infty,1}^{N}(\mathcal{L}),L_1^N\right)\ge \frac18.
\end{equation}
Consider the function $g:(0,1]\to \R$,
$$
g(x)=x\left(\log_2 e+\log_2 \frac1x\right). 
$$
It is elementary to check that $g$ is monotonely increasing. Moreover $g(x)\to 0$ as $x\to 0$.
Choose $0<c_2\le 1$ in such a way that
\begin{equation}
\label{BH6}
g(x)< \frac{c_1}{2} \quad( 0<x\le c_2).
\end{equation}
Now put
\begin{equation}
\label{BH3}
c=\min\left(\frac{c_1}{2\log_2\log_2 5 }, \frac{c_2}{2},\frac12\right)
\end{equation}
and assume 
\begin{equation}
\label{BH4}
n\le c N.
\end{equation}
If $n=0$, Corollary \ref{cor:B1} gives
\begin{equation}
\label{BH1}
e_0^\q(J_{\infty,1}^{N},\mathcal{B}(L_\infty^N))=\|J_{\infty,1}^{N}\|= 1.
\end{equation}
Hence we can suppose that $n\ge 1$, which, by (\ref{BH4}), implies $N\ge c^{-1}$. Consequently, from (\ref{BH3}),
\begin{equation}
\label{BH5}
\frac{\log_2\log_2 5}{N}\le \frac{c_1}{2}.
\end{equation}
Since by (\ref{BH3}) and (\ref{BH4}), $2n/N\le 2c\le c_2$, we get from (\ref{BH6})
\begin{equation}
\label{BB8}
\frac{2n}{N}\left(\log_2 e+ \log_2 \frac{N}{2n}\right)<\frac{c_1}{2},
\end{equation}
and therefore, with (\ref{BH5}),
\begin{equation}
\label{BB6}
\frac{\log_2 \log_2  5}{N}+\frac{2n}{N}\left(\log_2 e+ \log_2 \frac{N}{2n}\right)< c_1.
\end{equation}
This implies, using also (\ref{BF7}),
\begin{equation}
\label{BF9}
(\log_2 5) \left(\frac{eN}{2n}\right)^{2n}< 2^{c_1N}\le k+1.
\end{equation}
Since we have $k,n\in \N$ satisfying (\ref{BF9}), and moreover, by (\ref{BH3}) and (\ref{BH4}), $2n\le N$, we can use Proposition
 \ref{pro:B1} together with   (\ref{BG7}) and (\ref{BF8}) to conclude
\begin{equation*}
e_n^\q(J_{\infty,1}^{N},\mathcal{B}(L_\infty^N))
\ge \phi_k\left(J_{\infty,1}^{N}(\mathcal{L}),L_1^N\right)\ge \frac18.
\end{equation*}
\end{proof}

Using Proposition \ref{pro:B2} we can also remove the logarithmic factors in another lower bound -- for Sobolev embeddings 
$J_{pq}:W_p^r([0,1]^d)\to L_q([0,1]^d)$, see  \cite{Hei04b} for the notation and Proposition 2 of that paper for the 
previous result. The following can be derived from Proposition \ref{pro:B2} using the same argument as in \cite{Hei04b},
p.\ 43, relations (87) and (88).
\begin{corollary} Let $1\le p,q\le \infty$, $r,d\in \N$, and assume $\frac{r}{d}>\max\left(\frac1p,\frac2p-\frac2q\right)$.
Then there is a constant $c>0$ such that for all $n\in \N$
$$
e_n^\q(J_{pq},\mathcal{B}(W_p^r([0,1]^d)))\ge c n^{-r/d}.
$$
\end{corollary}

Furthermore, the lower bounds from \cite{Hei04a} were also used in \cite{Hei04c}, Proposition 3 and Corollary 3.
Using Proposition  \ref{pro:B2}, these results can be improved in the respective way, too. We omit the details.

Let us finally comment on lower bounds for power queries introduced in \cite{PW05}. An inspection of the proof 
of Lemma \ref{lem:B1} shows that the type of query is not used at all in the argument, so the statement also holds for power queries.
One part of the argument in both \cite{Bes04,Bes05} consists of proving that for a quantum algorithm with at most $n$ power queries and
for a suitable subset $F_0\subseteq F$, which can be 
identified with the interval $[0,1]$, the respective space $\mathcal{P}_{A,F_0}$
is contained in the (complex) linear span of functions $e^{2\pi i\alpha t}\;\,(t\in [0,1])$,  with frequencies $\alpha$ from a set of 
cardinality not greater than $c^n$ for some $c>0$, hence, $\dim \mathcal{P}_{A,F_0}\le 2c^n$. 
Moreover, also $S(F_0)$
can be identified with the unit interval. 
Now Lemma \ref{lem:B1} above 
directly yields the logarithmic lower bounds from \cite{Bes04,Bes05}, since the $k$-th inner entropy number of the unit interval
 is  $k^{-1}$.

\end{document}